\documentclass{osa-article}

\journal{osajournal}

\articletype{Research Article}

\usepackage{amsmath}
\usepackage{graphicx}

\newcommand{\sket}[1]{{\ensuremath{\lvert#1\rangle}}}
\newcommand{\lket}[1]{{\ensuremath{\left\lvert#1\right\rangle}}}
\newcommand{\ket}[1]{\if@display\lket{#1}\else\sket{#1}\fi}

\newcommand{\sbra}[1]{{\ensuremath{\langle#1\rvert}}}
\newcommand{\lbra}[1]{{\ensuremath{\left\langle#1\right\rvert}}}
\newcommand{\bra}[1]{\if@display\lbra{#1}\else\sbra{#1}\fi}

\newcommand{\sbraket}[2]{{\ensuremath{\langle#1\rvert#2\rangle}}}
\newcommand{\lbraket}[2]{{\ensuremath{\left\langle#1\!\left\rvert\vphantom{#1}#2\right.\!\right\rangle}}}
\newcommand{\braket}[2]{\if@display\lbraket{#1}{#2}\else\sbraket{#1}{#2}\fi}

\newcommand{\sketbra}[2]{{\ensuremath{\lvert #1\rangle\!\langle #2\rvert}}}
\newcommand{\lketbra}[2]{{\ensuremath{\left\lvert #1\right\rangle\!\!\left\langle #2\right\rvert}}}
\newcommand{\ketbra}[2]{\if@display\lketbra{#1}{#2}\else\sketbra{#1}{#2}\fi}

\begin{document}


\title{Heterodyne spectrometer sensitivity limit for quantum networking}

\author{Joseph C. Chapman\authormark{*} and Nicholas A. Peters}

\address{Quantum Information Science Section, Oak Ridge National Laboratory, Oak Ridge, TN 37831\footnote{Notice: This manuscript has been co-authored by UT-Battelle, LLC, under contract DE-AC05-00OR22725 with the US Department of Energy (DOE). The US government retains and the publisher, by accepting the article for publication, acknowledges that the US government retains a nonexclusive, paid-up, irrevocable, worldwide license to publish or reproduce the published form of this manuscript, or allow others to do so, for US government purposes. DOE will provide public access to these results of federally sponsored research in accordance with the DOE Public Access Plan (http://energy.gov/downloads/doe-public-access-plan).}}
\email{\authormark{*}chapmanjc@ornl.gov}


\begin{abstract}
Optical heterodyne detection-based spectrometers are attractive due to their relatively simple construction and ultra-high resolution.
Here we demonstrate a proof-of-principle single-mode optical-fiber-based heterodyne spectrometer which has picometer resolution and quantum-limited sensitivity around 1550 nm. Moreover, we report a generalized quantum limit of detecting broadband multi-spectral-temporal-mode light using heterodyne detection, which provides a sensitivity limit on a heterodyne detection-based optical spectrometer. We then compare this sensitivity limit to several spectrometer types and dim light sources of interest, such as, spontaneous parametric downconversion, Raman scattering, and spontaneous four-wave mixing. We calculate the heterodyne spectrometer is significantly less sensitive than a single-photon detector and unable to detect these dim light sources, except for the brightest and narrowest-bandwidth examples.
\end{abstract}


\section{Introduction}
In quantum networking, the spectra produced by the photon sources is critical to their performance. For quantum teleportation~\cite{quanttel} and entanglement swapping~\cite{entswappaper}, the photons entering the intermediate measurement system (Bell-state measurement~\cite{PhysRevA.51.R1727}) need to be indistinguishable~\cite{u2005generation}. To certify this, it often includes measuring the joint-spectral intensity of the photon pair source with a single-photon spectrometer~\cite{doi:10.1021/ac50039a022,doi:10.1063/1.121984,ToF1,ToF2,Chen:17,Chen:19,cheng2019broadband}. Moreover, adaptive bandwidth management has come to the forefront of current research as the community tries to circumvent the often-low emission rates of quantum technologies and provision the available photons to user needs~\cite{Lingaraju:21,Appas2021}. In this case, a single photon spectrometer is required to characterize the highly entangled joint-spectral intensity~\cite{PRXQuantum.2.040304}. These previous demonstrations have often necessitated use of expensive cryogenic single-photon detectors. It is therefore desirable to develop a single-photon spectrometer with high spectral resolution but without the use of cryogenic detectors.
 
A heterodyne spectrometer mixes the input signal with a tunable local oscillator and detects the strength of the intermediate frequency using a photodetector, e.g., a balanced photodiode pair followed by electrical amplification~\cite{1448067,blaney1975signal,menzies1976laser,parvitte2004infrared}. Over the years, heterodyne spectrometers have found many uses, including some of the earliest being characterization of atmospheric molecular composition~\cite{10.2307/1731444,Menzies:71}, astronomical observation~\cite{MUMMA1975}, and laser linewidth characterization~\cite{iet:/content/journals/10.1049/el_19800437}. More recently, heterodyne spectrometers are still primarily used for atmospheric molecular composition~\cite{Bomse:20,Deng:21,Sappey:21}.
Heterodyne spectrometers have the benefit of the frequency resolution being primarily limited only by the local oscillator bandwidth and stability, assuming the post-processing electronics have the required precision. Because of this, heterodyne spectrometers have been demonstrated with resolution $< 125$ MHz ($< 1$ pm at 1550 nm)~\cite{kovalyuk2017chip,s17020348,352070,Sonnabend:02,WIRTZ20022457}. The sensitivity of these heterodyne spectrometers on the other hand has not been as impressive, often not much below - 60 dBm~\cite{Furukawa:13,986811} when using photodiodes. Using superconducting-nanowire single-photon detectors, -126 dBm has been shown for narrowband signals~\cite{kovalyuk2017chip}. Others have shown that there are practical and theoretical limits to the sensitivity of heterodyne spectrometer and discussed them in the context of astronomical observation~\cite{blaney1975signal,AO1976} and laser side-band characterization~\cite{Boucher1993}.

In this work, we analyze the utility of heterodyne spectrometers for characterization of several light sources interesting to the growing quantum networking community. These very dim and often broadband light sources required specification of a more general sensitivity limit for heterodyne spectroscopy which we use in comparison with the brightness of several interesting light sources. In Sec. \ref{demo}, we demonstrate a proof-of-principle fiber-based heterodyne spectrometer which has picometer resolution and high sensitivity in the conventional optical communications band. In Sec. \ref{SA}, we generalize the sensitivity analysis of ~\cite{AO1976} with respect to the input spectra. Although our demonstration is in the infrared, our sensitivity analysis is applicable to any shot-noise limited heterodyne detector for any input spectra so is generally applicable to many use cases. We use our generalized analysis to compare the measured sensitivity for our heterodyne spectrometer with the generalized sensitivity limit, and with the measured sensitivity of a conventional grating-based spectrometer and a direct-detection single-photon-detector-based spectrometer. Finally, in Sec. \ref{MBDLS}, we compare the heterodyne-spectrometer-sensitivity limit with common single-photon sources, i.e., spontaneous parametric downconversion, spontaneous four-wave mixing, Raman scattering, and quantum dots.

\section{Fiber-based }
\begin{figure*}
    \centering
    \includegraphics[scale=0.6]{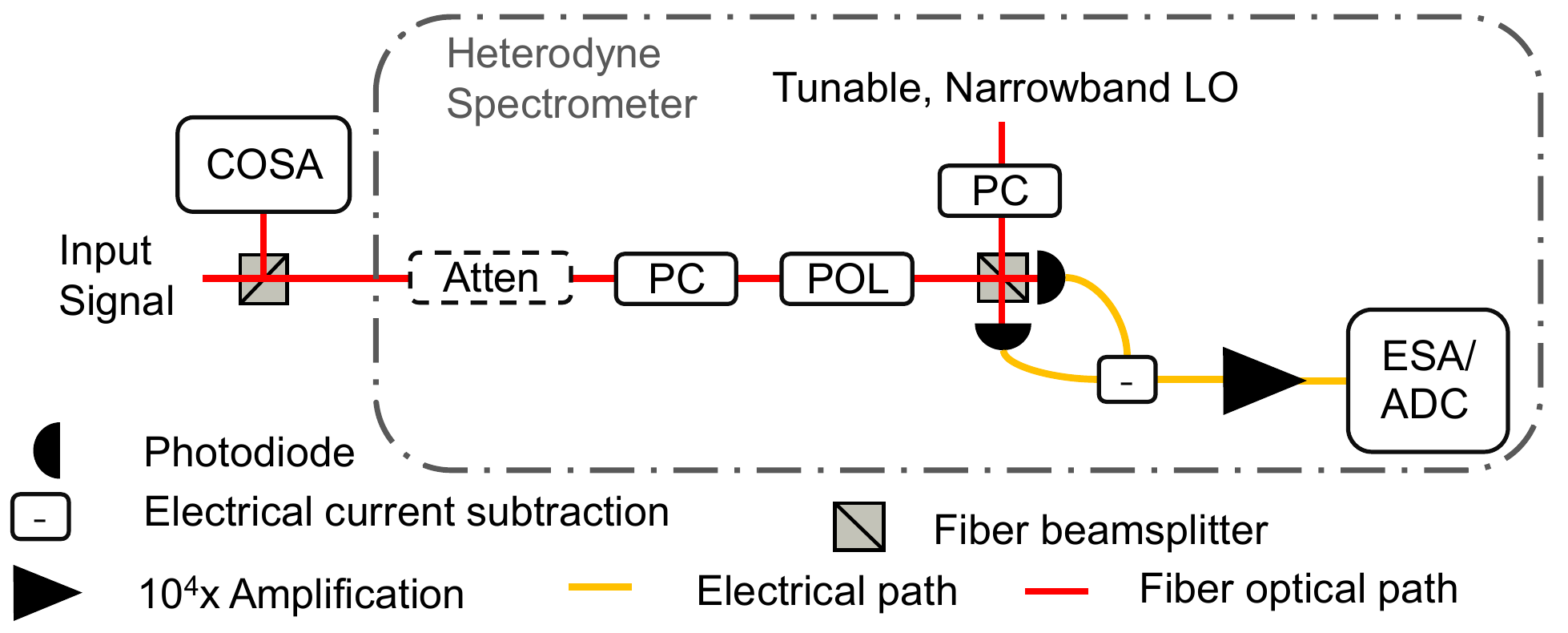}
    \caption{Spectrometer comparison experimental setup. Atten = Optional attenuator(s). COSA = Conventional optical spectrum analyzer. PC = Polarization controller. POL =  Polarizer. ESA/ADC = Electronic spectrum analyzer or analog-to-digital converter with microprocessor.}
    \label{fig:HTsetup}
\end{figure*}

Demonstration \label{demo}
In our simple  single-mode optical-fiber-based heterodyne spectrometer (see Fig. \ref{fig:HTsetup}), an input signal of any polarization, and potentially unknown spectra, enters the input port. It is attenuated, if necessary, and polarized. The polarization controller is used to optimize power through the polarizer. The input signal is mixed with the local oscillator on a 50/50 fiber beamsplitter. The mixing products are detected by an amplified balanced photodiode pair (Thorlabs PDB430C). For convenience, the amplified subtracted detector output goes to an electronic spectrum analyzer (ESA, specifically an Agilent N9000A CXA Signal Analyzer). Alternatively, if an ESA is unavailable or a smaller device is desired, an analog-to-digital converter (ADC) connected to a microprocessor could have been used~\cite{s17020348} for a more cost-effective and integrated analysis solution.

In this experiment, we use two different input signals. We use a steep-edge filtered (1 nm bandpass filter from the Finisar Waveshaper 1000A) amplified-spontaneous-emission (ASE) source (Pritel SCG-40) and a narrowband (100-kHz linewidth, broadened to 50-500 MHz) tunable laser (Hewlett Packard 8168C). To generate the tunable narrowband local oscillator (LO), we use a tunable continuous-wave (CW) external-cavity laser (Pure Photonics PPCL550) with an intrinsic linewidth of about 10 kHz, frequency noise-broadened to about 200 kHz for $>0.1$-ms integration times~\footnote{We estimate the broadened bandwidth using the technique outlined in \cite{DiDomenico:10}, which used the manufacturer's measurement of the frequency noise.}, and relative intensity noise < -120 dBc/Hz at 100 kHz. There is also an internal 900-Hz frequency dither on the LO, which further broadens the laser to about 100 MHz for $>1$-ms integration times. For simplicity of this demonstration, we utilize the manufacturer's wavelength calibration of the LO. Others have already demonstrated real-time  wavelength calibration that could be used instead~\cite{s17020348}. For comparison, we also record spectra on a conventional grating optical spectrum analyzer (Yokogawa AQ6370B) which has up to 20 pm resolution and -90 dBm sensitivity, implying a power-spectral-density sensitivity of -90 dBm/20 pm for signals of bandwidth $\geq$ 20 pm. The input signal was split with a 50/50 fiber beamsplitter between the Yokogawa optical spectrum analyzer (OSA) and the heterodyne spectrometer.

\begin{figure*}
    \centering
    \includegraphics[scale=0.45]{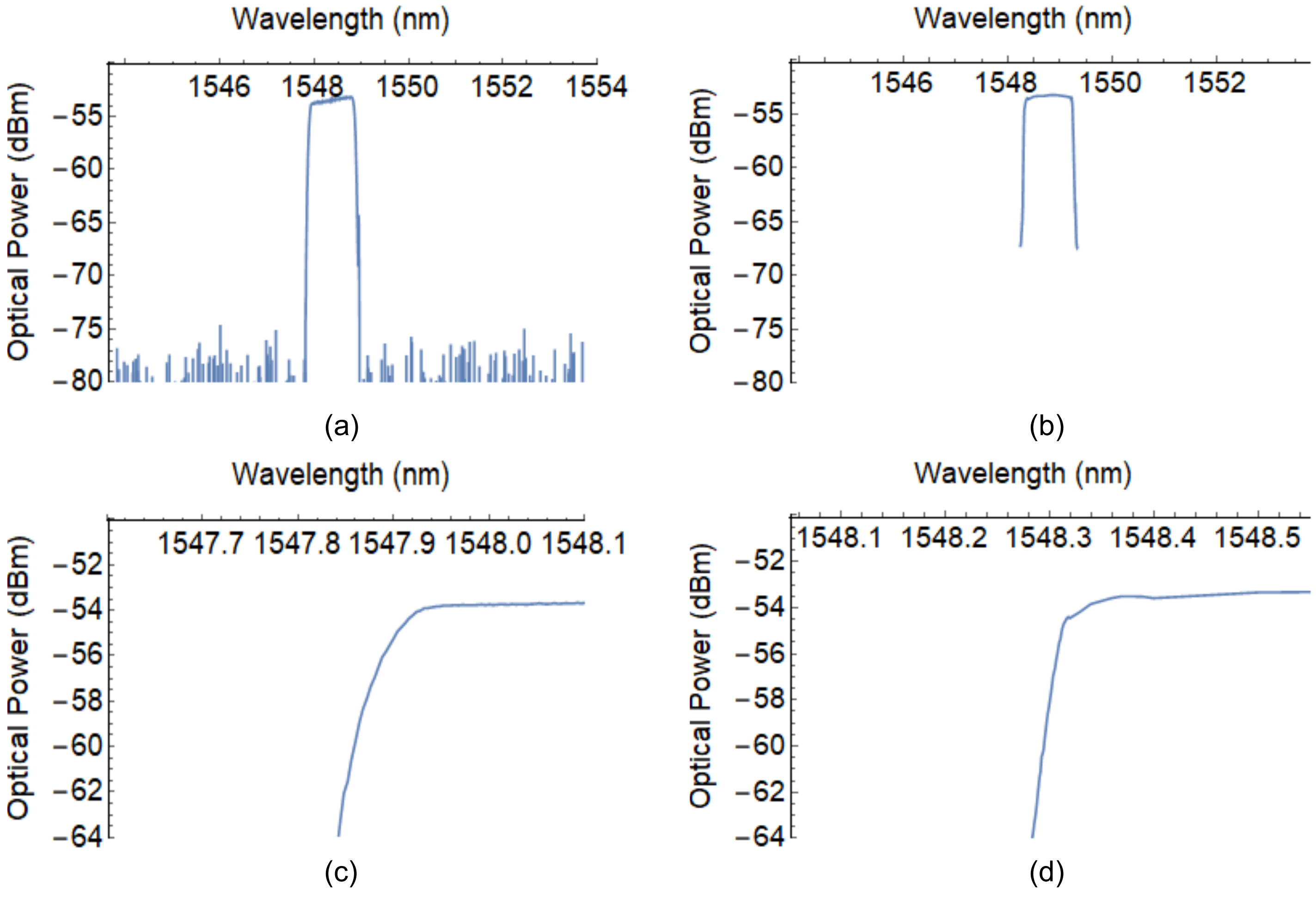}
    \caption{Spectrometer comparison with 1-nm input signal. (a) Yokogawa spectrometer output using best resolution (20 pm). Note: the absolute wavelength calibration of the Yokogawa OSA differs from our other instruments by about 0.4 nm. (b) Heterodyne spectrometer output. (c) Magnified top left corner of (a). (d) Magnified top left corner of (b).}
    \label{fig:HTdata1nmfilt}
\end{figure*}

We measure the spectra of the steep-edge filtered ASE source using the Yokogawa OSA and our heterodyne spectrometer (see Fig. \ref{fig:HTdata1nmfilt}). Due to the < 10 GHz minimum bandwidth of the programmable filter used, we expect very steep edges. Fig. \ref{fig:HTdata1nmfilt}a is the output of the Yokogawa OSA on the highest resolution setting (20 pm) and Fig. \ref{fig:HTdata1nmfilt}b is the output of the heterodyne spectrometer. Notice the rounded edge of the magnified graph in Fig. \ref{fig:HTdata1nmfilt}c compared to the steeper edge and increased detail in Fig. \ref{fig:HTdata1nmfilt}d; this indicates the heterodyne spectrometer has a better resolution.

\begin{figure*}
    \centering
    \includegraphics[scale=0.45]{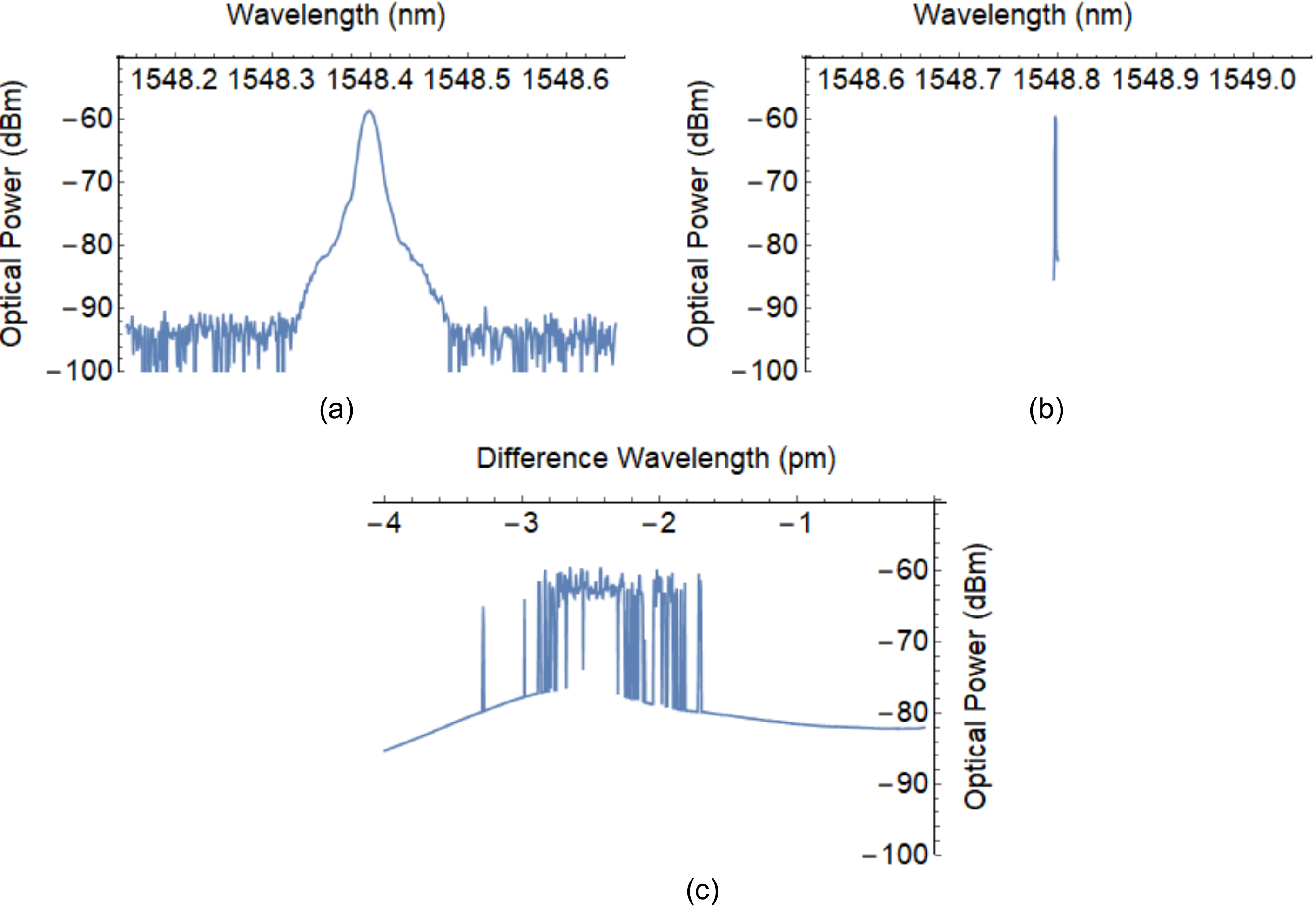}
    \caption{Spectrometer comparison with narrowband laser input. (a) Yokogawa spectrometer output using best resolution (20 pm). Note: the absolute wavelength calibration of the Yokogawa OSA differs from our other instruments by about 0.4 nm. (b) Heterodyne spectrometer output. Data was taken with a ESA video bandwidth of 10 Hz which partially integrates over LO frequency dither. (c) Magnified version of (b). $\lambda_{\text{LO}}=1548.800$ nm.}
    \label{fig:HTdataHPlaser}
\end{figure*}

In Fig. \ref{fig:HTdataHPlaser}, we compare the resolution more directly and can estimate it straight from the data. Here the input signal is a narrowband tunable laser, which is broadened, presumably by internal frequency dithering. In Fig. \ref{fig:HTdataHPlaser}a, we see a peak produced by the input signal being detected by the Yokogawa spectrometer. In Fig. \ref{fig:HTdataHPlaser}b, on the same scale, we see a much narrower peak produced by the same input signal being detected by the heterodyne spectrometer. A magnified graph (Fig. \ref{fig:HTdataHPlaser}c) reveals the input signal linewidth convolved with the LO linewidth and the dithering of the two lasers. From this we can see the true laser linewidths of the LO and input signal are each much less than 1 pm.

We see in Fig. \ref{fig:HTdataHPlaser}c artifacts of the LO operating mode which uses a frequency dither that spanned about 100 MHz, at a rate of about 900 Hz. The narrowband input signal was also broadened, likely via dithering as well. The video bandwidth during this data collection partially integrated the dither, resulting in many apparent peaks. With further development and automation, a frequency scan of the LO can be used to avoid an LO dither and the scan can be synchronized with the detector measurement by an ADC using a control microprocessor~\cite{s17020348}. Using a frequency scan, the spectrometer resolution can be improved significantly and is then limited by the laser linewidth, scan speed, and measurement integration time. Using these techniques near 1550 nm, resolution down to 6 MHz has been demonstrated~\cite{s17020348}.

\begin{figure*}
    \centering
    \includegraphics[scale=0.45]{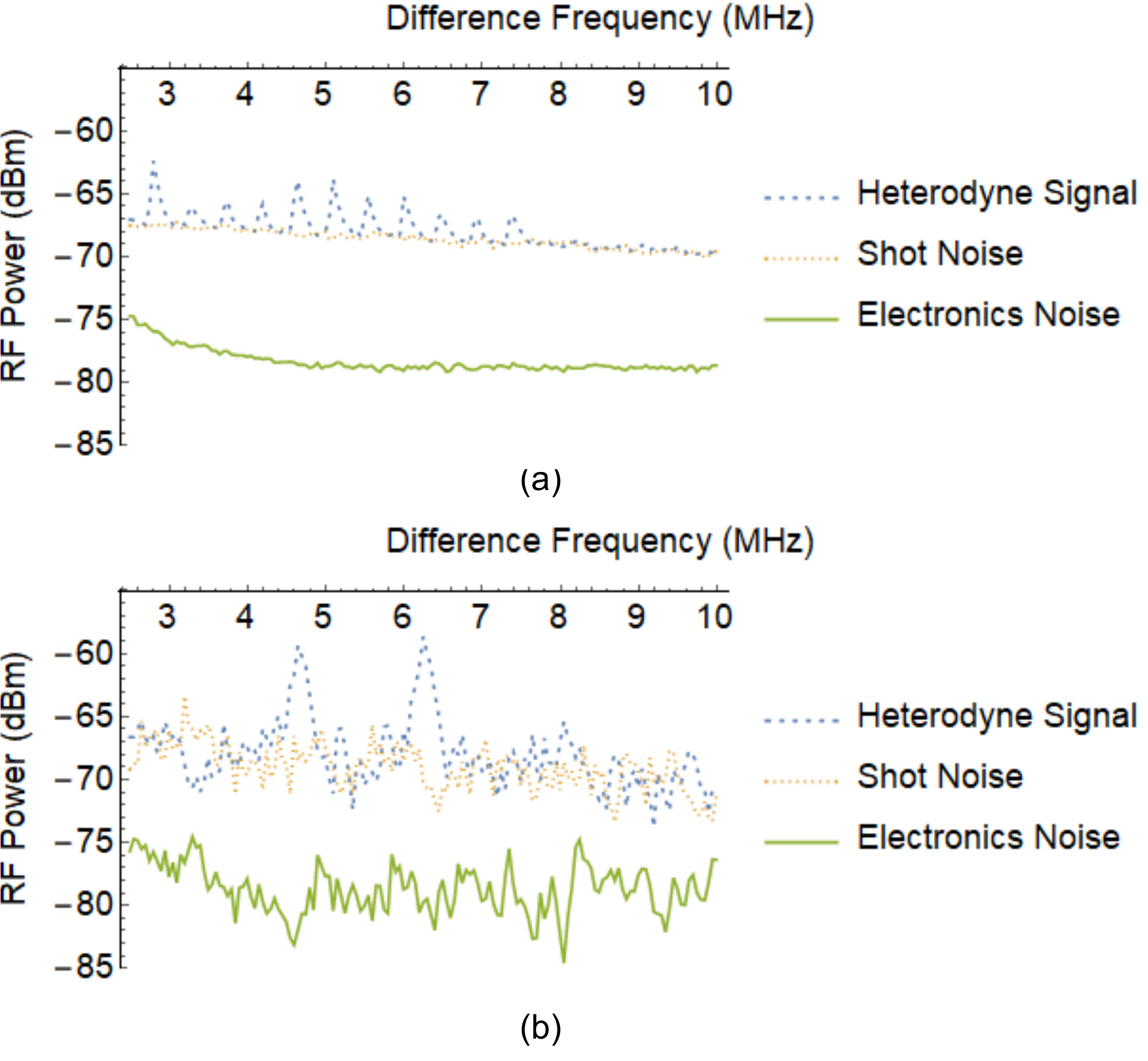}
    \caption{Heterodyne spectrometer output at the sensitivity level with $\lambda_{\text{LO}}=1548.809$ nm. (a) The input signal is the narrowband laser, attenuated to -89 dBm (optical power). The ESA video bandwidth was 1 kHz, and the resolution bandwidth was 1 MHz. RF = radio frequency. (b) Using an ESA 100-kHz video bandwidth, we remeasure the heterodyne signal from the same input signal as (a).}
    \label{fig:HTsensdata}
\end{figure*}

Furthermore, for amplification, our Thorlabs PDB430C uses two gain stages of $10^3$ and 10, via two Analog Devices OPA847 amplifiers. We modified another PDB430C we have to a single gain stage of $10^4$ using one Analog Devices OPA657 and an output low-pass filter with a corner frequency of about 10 MHz. These modifications significantly reduce the bandwidth from about 350 MHz to about 10 MHz, but they also reduce the electronics noise by about 8 dB. With these modifications, we improve on Fig. \ref{fig:HTdataHPlaser}b; the heterodyne spectrometer now has a sensitivity of about -89 dBm (see Fig. \ref{fig:HTsensdata}a), roughly matching the Yokogawa OSA, and to our knowledge, now exceed the power sensitivity of all other demonstrated heterodyne spectrometers using photodiodes which published data providing absolute input signal power of the noise floor. Unfortunately, most previous heterodyne spectrometer demonstrations only published normalized power measurements.

While we believe we have developed the most sensitive hetrodyne spectrometer with photodiodes to date, it is natural to consider if the sensitivity may be further improved.  In Fig. \ref{fig:HTsensdata}b, we use a higher video bandwidth on the ESA which shows higher peaks for the same optical input power as Fig. \ref{fig:HTsensdata}a. This leads us to conclude that without frequency dithered lasers (which the LO and signal were), the sensitivity can be further improved.  In the next section, we consider the practical sensitivity limits for a heterodyne spectrometer. Accordingly, we compare the measured sensitivity of our spectrometer to the sensitivity limit described in Sec. \ref{SA} and show what the sensitivity could be without laser frequency dithering.

\section{Sensitivity Analysis}
\label{SA}
Previous work shows that heterodyne detection~\cite{Ref1AO1976}, and even heterodyne spectrometers~\cite{AO1976}, when LO shot noise dominates all other sources of noise,  can reach the quantum detection limit ($P_{min}$), i.e., detection of one photon (with energy $h \nu$) within the system resolution time $\Delta \nu^{-1}$ (sec),
\begin{equation}
    P_{min} = h \nu \Delta \nu,\label{Pmin}
\end{equation}
assuming the detector has unit quantum efficiency, where $h$ is Planck's constant and $\nu$ is the frequency of the light.

This detection limit and our analysis assume there is mutual spatial coherence~\cite{wolf2007introduction} between the signal and local oscillator. To satisfy this, we assume in our analysis that both the signal and local oscillator are combined into the same single spatial mode before detection, e.g., through careful optical alignment of single-spatial-mode beams or coupling into single-mode fiber-optic cable. Steady-state temporal coherence, as normally required for conventional interferometry is not required here.

More crucially, in Eq. (\ref{Pmin}) there is an implicit assumption that there is just a single spectral-temporal mode under consideration too. This may be true for narrow molecular transition lines, for which previous analyses were developed~\cite{AO1976, Boucher1993}, but in general, light sources emit into many spectral-temporal modes. When that is the case, the above analysis implies, and we state explicitly, that detecting a single photon per mode within the resolution time is the more general limit for heterodyne detection and therefore, also heterodyne spectrometers. For a single polarization, the number of spectral-temporal modes within a certain bandwidth $\Delta \nu$ and integration time $\Delta t$ is $N=\Delta \nu \Delta t=c\Delta \lambda \Delta t/\lambda^2$ ~\cite{Qi2017RSI}. For certain pulsed experiments, $\Delta t$ is instead proportional to the pulse width~\cite{PhysRevLett.121.083602}.  For a common integration time of 1 second, and a 1-kHz bandwidth, that implies one thousand spectral-temporal modes. Even narrow-linewidth sources ($\sim$ kHz) emit into many spectral-temporal modes, let alone broadband sources, so this is an important consideration.

For a heterodyne spectrometer, the average detected photon number per mode is

\begin{equation}
    \langle n(\nu_{\text{in}}) \rangle  = \frac{S_{\text{in}}(\nu_{\text{in}}) \eta }{h \nu_{\text{in}}} \label{nin}
\end{equation}
where the input signal power-spectral density is $S_{\text{in}}(\nu_{\text{in}})$ and the detection efficiency is $\eta$. The input signal power-spectral density is equivalent to the average energy detected per temporal-spectral mode, using dimensional analysis. Thus, Eq. (\ref{nin}) is the average energy detected per temporal-spectral mode divided by the energy per photon ($h \nu_{\text{in}}$) which equals the average detected photon number per mode. Shot noise contributes on average one photon per mode~\cite{AO1976,Qi:20}, thus, the signal-to-noise ratio (SNR) is $
    \text{SNR}(\nu_{\text{in}})=\frac{ \langle n(\nu_{\text{in}}) \rangle}{1}= \langle n(\nu_{\text{in}}) \rangle$.

The sensitivity limit is defined above as one detected photon per mode, equates to a measured noise variance 3-dB higher than the shot noise, when the input noise quadratures are equal. To derive this we use the relation from \cite{Qi:20} which equates average photon number per mode to measured noise variance.
\begin{align}
    \langle n \rangle  &=  \langle Z \rangle - 1 \label{nz}\\
    &= \langle X^2 \rangle + \langle P^2 \rangle - 1\\
     &= 2 \langle \Delta X^2 \rangle -1, \label{ndelX}
\end{align}
where we use the definition $Z=X^2+P^2$, and the assumptions  $\langle \Delta P^2 \rangle=\langle \Delta X^2 \rangle=\langle X^2 \rangle -\langle X \rangle^2$ and $\langle X \rangle=\langle P \rangle=0$. These are valid assumptions for symmetric phase-space distributions when averaging over the phase between the LO and input signal. Using Eq. (\ref{ndelX}), when $\langle n \rangle  = 0$, $\langle \Delta X^2 \rangle=1/2$ (shot-noise variance). Moreover, when $\langle n \rangle  = 1$, $\langle \Delta X^2 \rangle=1$, which is twice (3 dB) the shot-noise variance. Therefore, when the measured noise variance is 3 dB above the shot-noise variance, there is one photon per mode on average. Similarly, the noise-equivalent power of a coherent detector is on average one photon per mode, when the noise is dominated by LO shot noise and the $-1$ in Eq. (\ref{nz})-(\ref{ndelX}) is interpreted as the shot noise contribution~\cite{Qi:20}.

Using the same filtered ASE source from Fig. \ref{fig:HTdata1nmfilt}b, we attenuate its input power into the heterodyne spectrometer until we measure an radio-frequency (RF) noise power of about -65.5 dBm at 6 MHz, which is 3-dB greater than the shot-noise (-68.5 dBm, which is 10 dB greater than the electronics noise using the modified detector) measured on the ESA. These measurements were taken with an integration time of 0.00012 s (one point of a 1001-point sweep of the ESA taking 0.12 s) with an ESA resolution bandwidth of 1 MHz. To measure the optical input power of the filtered ASE source sent into the heterodyne spectrometer when it produced a heterodyne signal 3 dB above shot noise, we use the Yokogawa OSA as a power meter and measured an optical power-spectral-density of $S_{\text{in}}=$-64 dBm/20 pm, near $\nu_{\text{LO}}$. Using Eq. (\ref{nin}), this implies we need about 1.25 input photons per mode from the filtered ASE source to have one detected photon per mode. Thus, the heterodyne spectrometer has a measured sensitivity of -64 dBm/20 pm. The necessary input photons per mode are greater than one because of electronics noise, loss, and imperfect detection efficiency. This shows our spectrometer is very near the quantum limit for shot-noise-limited detection sensitivity.

Using the narrowband laser (from Fig. \ref{fig:HTdataHPlaser}b), we attenuate its input power into the heterodyne spectrometer until we measure a signal about 3 dB above shot noise (see Fig. \ref{fig:HTsensdata}a). At this optical power level, we use the Yokogawa OSA as a power meter and measure an optical power of -89 dBm. The laser is specified to have a 100-kHz linewidth; thus, the actual power-spectral density sensitivity measured is -89 dBm/0.8 fm = -45 dBm/20 pm, assuming a top-hat spectral shape with a width of 100 kHz. This means that for 1 detected photon per mode, about 100 input photons per mode are needed, much greater than one. This sensitivity is primarily due to the internal laser frequency dithering of both the LO and the narrowband signal laser, and other noise broadening of the laser linewidths. Accordingly, there is not always an intermediate frequency produced at 6 MHz (our detection frequency), or even in the detection bandwidth (see Fig. \ref{fig:HTsensdata}b). If the lasers were not dithered or broadened, these measurements imply a possible power-spectral-density sensitivity of about -109 dBm/0.8 fm = -65 dBm/20 pm, which is roughly equal to the directly measured sensitivity using the filtered ASE source. The sensitivity for the heterodyne spectrometer is thus consistent, namely, one detected photon per spectral-temporal mode, regardless of the input light source.

In contrast, the Yokogawa OSA measures power-spectral density with noise that is fixed for a given spectral resolution. Importantly, it does not have any noise contribution from LO shot noise (since there is no LO used) so it is not limited by Eq. \ref{Pmin}. At the minimum resolution, the power-spectral-density sensitivity is -90 dBm/20 pm. That is equivalent to 0.003 photons per spectral-temporal mode. Nevertheless, this noise is fixed. If the signal occupies less than a 20 pm bandwidth, there is still the same amount of noise when measured with the Yokogawa OSA, which will effectively increase the input photons required per spectral-temporal mode. In that situation, it is possible for the heterodyne spectrometer to have better power sensitivity, assuming the LO linewidth is much less than 20 pm, as it is in our experiment. For example, it has been shown that a heterodyne spectrometer measured the resonance fluorescence of a single ion~\cite{doi:10.1080/09500349708231861}. These conclusions agree with a similar analysis which compared direct detection and heterodyne detection for astronomical sources of varying bandwidth~\cite{blaney1975signal}.

Let us now compare the sensitivity of a heterodyne spectrometer and the Yokogawa OSA to one based on a single-photon detector~\cite{cheng2019broadband}. Superconducting-nanowire single-photon detectors (SNSPD) have exceptionally low noise characteristics in the near infrared~\cite{yamashita2011origin,shibata2015ultimate}. For a SNSPD, sensitive around 1550 nm, coupled to single-mode fiber, a typical dark noise count rate is about 100 counts/s~\cite{yamashita2011origin,doi:10.1063/5.0006221}, depending on the temperature and bias current. This dark count rate is independent of any optical filter in front of the detector. For an even comparison, let us say there is a 20-pm wide tunable filter in front of the SNSPD. This notional configuration yields $4\times10^{-8}$ noise counts per mode. Thus, for dim broadband signals considered here, the filter and SNSPD is the better choice because it has lower noise. If the noise counts per mode using a filter and an SNSPD exceeds one, then it is more advantageous to use a heterodyne spectrometer. Interestingly, there was a demonstration with SNSPDs integrated into a heterodyne spectrometer~\cite{kovalyuk2017chip}. This device showed detection at low light levels (about 1000 photons/sec) for a very narrowband light source (about 1 kHz), which agrees with the shot-noise detection limit discussed here. Overall, these results agree with the consensus in the astronomical community that single-photon detectors are preferred for viewing dim astronomical objects~\cite{NAP26141,Nightingale1990}.

\section{Modal Brightness of Dim Light Sources}
\label{MBDLS}
Now we turn our attention to several light sources, which are commonly detected by single-photon detectors and are of great importance to the quantum networking community, to see if a heterodyne spectrometer could be used for their characterization.

Spontaneous parametric downconversion (SPDC)~\cite{SPDC} is a spectrally rich and diverse process, heavily used in quantum communications experiments. SPDC pair generation is highest most often in waveguides due to better mode overlap between the signal, idler, and pump and increased SPDC spectral density inside the waveguide~\cite{Fiorentino:07}. Waveguide pair generation rates of about $R_p=3\times10^8$ pairs/s per mW of pump in a 1-nm bandwidth have been measured for type-I SPDC processes in lithium niobate~\cite{Clausen_2014}. Here we are interested in calculating how many photons are generated into a single spectral-temporal mode. For $\lambda=1550$ nm, $\Delta \lambda=1$ nm, and $\Delta t=1$ s, there are about $N=10^{11}$ spectral-temporal modes. For a 1-mW laser-pumped SPDC source, that gives on average $3\times10^8/10^{11}=0.003$ photons per mode, resulting in an SNR much less than one, which means the signal is not practically detectable with a heterodyne spectrometer. A practical application of heterodyne spectrometry here would require scanning quickly over many frequency settings to capture the broadband spectra, leaving little time to integrate for each data point. It is not even likely that integration would help much for this application anyway, as it would help define the average noise amount better, but \emph{not} remove it to reveal the true signal and improve the dynamic range. 

There has been development of factorable joint-spectral intensity SPDC sources, employing pulsed pump lasers~\cite{PhysRevLett.100.133601,Halder:09,Harder:13,Meyer-Scott:18}, which is important for entanglement swapping and other experiments involving multi-photon interference~\cite{PhysRevLett.59.2044,PhysRevLett.84.5304,PhysRevA.77.022312}. The brightness of these sources is still much less than one photon pair per pump pulse, so that even if an LO sharing the same spectral-temporal mode of the SPDC were used, it would still not be easily detectable with a heterodyne spectrometer. Furthermore, since the SPDC and LO would be in the same mode, the spectrometer resolution would be equal to the convolution of spectral width of the SPDC spectrum.

Raman scattering is commonly produced in fiber-optic cables by bright laser light (such as the pulses which carry data in fiber-optic networks) inelastically scattering with the fiber itself~\cite{SRS}. Raman scattering is a pervasive noise source that often degrades the results of demonstrations where quantum and classical signals coexist together in the same fiber~\cite{Peters_2009,PhysRevX.2.041010,Niu:18,Tessinari:21}. The bandwidth of Raman scattering in optical fibers is usually several Tera-Hertz and the scattering cross-section ($\rho(\lambda)$) is on the order of $10^{-9}$ nm$^{-1}$ km$^{-1}$~\cite{10.1117/12.2306875}. For a $P_0=1$ mW laser going down a $L=25$ km standard single-mode fiber (with attenuation per unit length $\alpha=0.2$ dB/km), that produces at the output of the fiber~\cite{10.1117/12.2306875}

\begin{align}
    P_{\text{SRS}} &= P_0 L \rho(\lambda)10^{\alpha L/10}=8\times10^{-11}\text{W}/\text{nm}\\
    &=6\times10^8 (\text{1550-nm photons/s/nm}).
\end{align}
For an 1-sec integration time and a 1-nm bandwidth at 1550 nm there are about $N=10^{11}$ spectral-temporal modes. In that case, the number of SRS photons per mode is about $6\times10^8/10^{11}=0.006$, resulting in an SNR much less than one, which means the signal is not practically detectable with a heterodyne spectrometer. Using a pulsed pump with a high peak power ($> 1$ W) can greatly increase the Raman scattered photons within the pulse and the scattered photons may be able to be seen on a heterodyne spectrometer with the right LO (which would also likely need to be pulsed).

Spontaneous four-wave mixing (SFWM) is a process where two degenerate pump photons are converted into a signal-idler pair of photons~\cite{horowicz1987generation, pinard1989self, grandclement1989four, vallet1990generation}. SFWM is a competing process to SPDC for the generation of photon pairs. It can occur in fiber-optic cables rather efficiently due to long interaction lengths and small mode volumes~\cite{Wang_2001}. This process occurs most efficiently when there is optimal phase matching which happens when the pump is located at the zero-dispersion wavelength of the fiber~\cite{mechels1997accurate}. The average number of photons per mode for the signal and idler beams is $|\gamma P_0 L|^2$, where $\gamma$ is the non-linear coefficient of the fiber (often having units of W$^{-1}$ km$^{-1}$), $P_0$ is the pump power, and $L$ is the fiber length~\cite{Wang_2001}. Highly non-linear fibers can have $\gamma = 10$ W$^{-1}$ km$^{-1}$ and lengths commonly less than 1 km. With a pump power of 1 mW, the average number of photons per mode is $10^{-4}$. Clearly, this source of light is not bright enough to be practically detected by a heterodyne spectrometer either.

Finally, there is current development of deterministic single-photon sources using quantum dots~\cite{senellart2017high}. These narrow-linewidth emitters, at first, appear to be a good candidate for characterization with a heterodyne spectrometer, but they will not be bright enough since they are emitting equal to or less than one photon per spectral-temporal mode, as ``single'' photon sources. If the quantum dot sources were operated with higher brightness, then spectral characterization via a heterodyne spectrometer may be possible.

\section{Conclusion}
We have demonstrated a proof-of-principle heterodyne spectrometer which has a 20-times better wavelength resolution than a conventional low-noise grating-based spectrometer, with the potential for 200-times better resolution. Furthermore, the heterodyne spectrometer sensitivity can be much better than the conventional spectrometer for signals narrower than the conventional spectrometer's minimum resolution, otherwise the heterodyne spectrometer sensitivity is worse. Moreover, we calculate that, due to LO shot noise, heterodyne-detection-based spectrometers have fundamental sensitivity limitations that are significantly higher than that of a single-photon detector. Finally, we analyze the brightness of several light sources of interest to quantum networking and compare them with the heterodyne spectrometer sensitivity limit. From this comparison, it is now clear that the heterodyne spectrometer is not sensitive enough to measure spectra of many light sources of interest in quantum networking, specifically those of broad bandwidth that have on average much less than one photon per mode yet, it is sensitive enough for brighter narrow linewidth sources.

\begin{backmatter}

\bmsection{Acknowledgments}
The authors acknowledge Bing Qi for his advice on accurately counting spectral-temporal modes.
This work was performed at Oak Ridge National Laboratory, operated by UT-Battelle for the U.S. Department of Energy under contract no. DE-AC05-00OR22725.
Funding was provided by the U.S. Department of Energy, Office of Science, Office of Advanced Scientific Computing Research, through the Transparent Optical Quantum Networks for Distributed Science Program (Field Work Proposal ERKJ355).\\

\bmsection{Disclosures}
The authors declare no conflicts of interest.

\bmsection{Data Availability Statement}
Data underlying the results presented in this paper are not publicly available at this time but may be obtained from the authors upon reasonable request.
\end{backmatter}


\end{document}